# FRACTURE NUMERIQUE CHEZ LES SENIORS DU 4ème AGE.

## *Observation d'une acculturation technique*

CHRISTINE MICHEL
UNIVERSITE DE LYON/INSA-LYON/LIESP
Christine.Michel@insa-lyon.fr

MARC-ERIC BOBILLIER-CHAUMON
UNIVERSITE DE LYON/Université Lyon2/GRePS
Marc-Eric.Bobillier-Chaumon@univ-lyon2.fr

FRANCK TARPIN-BERNARD
UNIVERSITE DE LYON/INSA-LYON/LIESP
SBT, VILLEURBANNE
Franck.Tarpin-Bernard@insa-lyon.fr

**Contexte et enjeu**

En vieillissant, les personnes âgées (PA) accumulent les « handicaps » : sociaux, physiques, psychologiques ou cognitifs (Gorgeon et Léridon, 2001, Plonton, 2003). Il s'agit d'abord d'un déclin cognitif (avec une réduction des possibilités d'adaptation, des désapprentissages, de la démotivation, des difficultés de mémorisation…) et de dégradations psychologiques importantes (marquées par une plus grande vulnérabilité psychologique, l'absence de nouveaux investissements, une atteinte de l'estime de soi, la dépression ….). L'effritement de l'identité et du lien social est également spécifique de cette génération (Meire, 1992, David & Starzec, 1996). Avec le grand âge, on observe ainsi un repli de la personne sur le domicile et un affaiblissement significatif de ses rôles sociaux et familiaux, une vie par procuration, des conduites régressives (alimentation, hygiène, usages sociaux), une perte de but et d'identité conduisant à un état d'anomie (Atchley, 1980). Ce désengagement social s'exprime notamment par la diminution du niveau d'interaction sociale tant par la fréquentation que par le degré d'implication. Ainsi près de 65% des plus de 75 ans vivent une situation d'isolement, c'est-à-dire qu'ils n'ont ni sorties, ni relations, ni contacts téléphoniques avec des tiers (famille, amis…) (David & Starzec, 1996). Diverses recherches se sont ainsi développées pour cerner les attentes et besoins des PA et voir le bénéfice que pouvait produire les technologies sur leurs conditions de vie. Ces approches regroupées sous le terme de « gérontechnologie » s'appuient sur des dispositifs technologiques qui peuvent aider les PA à identifier et ralentir les effets de l'âge sur les systèmes neuronaux ou locomoteurs (Micera, Bonato, et Tamura, 2008). En réduisant les diverses dégradation sensori-motrices ou cognitives, elles peuvent améliorer la qualité de vie et la capacité que peuvent avoir les PA à participer aux activités journalières, à favoriser le maintien à domicile (en réduisant les séjours dans les résidences spécialisées (hôpitaux, EHPAD[1], maison de retraite…) et aussi accroitre leur autonomie. A titre d'exemples, Furness (2007) présente un sol qui est capable de suivre les mouvements des seniors et donner l'alerte en cas de chute, ou encore un ordinateur intégré à une armoire à pharmacie qui vérifie la prise de médicaments et alerte en cas d'erreur. Parmi les dispositifs TIC on peut citer les portails web comme *Care Online* (Osman, 2005) ou *NIH Senior Health* (Morrell, 2005), les outils de messageries spécifiquement adaptées (Dickinson *et al.,* 2005) ou les jeux de stimulations cognitives (Activital, 2008).

On pourrait dire, a priori que l'offre de service numérique aux PA participe à une justice sociale. Rawls, philosophe américain propose une théorie de la justice sociale fondée sur deux principes qui sont présentés de manière synthétique par Lecomte (Lecomte, 2000). Le premier principe, dit *principe d'égale liberté* est prioritaire et vise à garantir des libertés et des droits égaux pour tous. Le second principe prône *l'égalité des chances* dans l'accès à diverses fonctions et positions sociales tout en respectant les *différences entre les individus*, en stipulant en particulier que les inégalités socio-économiques sont justes, si et seulement si elles produisent, en compensation, des avantages sur les plus défavorisés. En considérant cette définition, nous allons tenter de voir si c'est effectivement le cas pour les PA. Nous présenterons les grandes tendances de l'offre numérique aux PA, l'usage qui en est fait et le bénéfice, en termes de qualité de vie, qu'elle leur apporte. Pour mieux définir l'adoption de la technologie, l'usage et ainsi mieux interpréter le bénéfice, nous proposons d'identifier les dynamiques d'acceptation technologique des PA selon le TAM (*Technology Acceptance Model*) de Davis adapté (Hamner et Qazi, 2008) au travers d'un état de l'art et d'une enquête, qualitative et quantitative, réalisée auprès d'une population de PA du 4eme âge.

---

[1] EHPAD : Etablissement d'Hébergement pour les Personnes Agées Dépendantes.

**PA et justice sociale**

Sur les trois principes de Rawls on peut dire que la justice sociale n'est que rarement observée concernant l'accès et l'usage des TIC par les PA. En effet, comme le souligne le rapport « France numérique 2012 » (Besson, 2008): *« Même si environ un million de seniors supplémentaire ont acquis une connexion Internet en 2007 (vs 2006), il est nécessaire de réduire la fracture numérique sur cette cible, car plus de 5,7 millions de seniors restent "e-exclus" encore aujourd'hui en France ».* Ce que confirme d'ailleurs l'enquête menée par le Credoc (Bigot et Croutte, 2007) qui montre que l'âge est déterminant dans l'achat des environnements TIC et dans leur utilisation quotidienne. *« Les sexagénaires comptent dans leurs rangs deux fois moins d'utilisateurs quotidiens que la moyenne (24% contre 49%). Passés 70 ans, le lien avec la machine est des plus ténus. ».* En effet, 86% des personnes de plus de 70 ans n'ont pas d'ordinateur et parmi celles qui en ont un, seul 46% se connectent tous les jours contre 30% qui se connectent moins que deux fois par semaine voire jamais. Mais, comme pour d'autres populations, plus que les questions d'accès, ce sont des questions d'utilisation et de qualité de l'utilisation (Camacho, 2005) qui accroissent la fracture numérique. Camacho mentionne que : *« Les fractures numériques résultent des possibilités ou des difficultés pour les groupes sociaux de mettre à profit, collectivement, les technologies de l'information et de la communication afin de transformer la réalité dans laquelle ils évoluent et d'améliorer les conditions de vie de leurs membres. »* De plus, s'il existe quelques dispositifs spécifiquement développés pour les PA comme nous l'avons vu ci-dessus, il faut bien savoir que la plupart des services (administratifs, bancaires, d'achat, médicaux…) délivrés par des supports virtuels et médiatisés, ainsi que leurs modalités d'interaction ou leurs type de ressources proposées ne sont pas spécifiquement adaptées à ce type de public (Godfrey et Johnson, 2008). Sachant que les PA sont particulièrement intéressés pour avoir des informations relatives à la santé et la médecine, Morrell (Morell, 2005) mentionne l'étude de Becker qui a analysé 125 sites web de ressources de santé. Ses résultats montrent des défauts d'utilisabilité pour les PA : 93% utilisent de petites polices de caractères, 24% nécessitent une utilisation complexe de la souris pour naviguer, 30% présentent un contenu informationnel compréhensible uniquement pour les personnes ayant fait des études supérieures. Enfin, si la prochaine génération semble être davantage sensibilisée aux technologies, le rythme effréné des changements qui caractérisent les innovations techniques risque de perpétuer le problème de l'adaptation au-delà des générations actuelles (Marquié et Baccarat, 2001). Les TIC, mal adaptées ou mal proposées, risquent d'accentuer l'exclusion sociale des PA et la fracture numérique alors même qu'elles laissaient entrevoir de formidables opportunités pour l'amélioration de leur qualité de vie.

Les gouvernements ont intégré ce fait dans leurs investissements puisque, si le financement de l'infrastructure reste prépondérant dans les plans de réduction de la fracture numérique, nous pouvons observer qu'ils sont complétés par d'autres directement orientés vers le soutien à l'usage. A titre d'exemple, dans le cas des seniors, citons l'action 27 du plan de développement de l'économie numérique (Besson, 2008) *« Action n°27 : Favoriser l'usage du numérique par les seniors. Lancer en 2009 une expérimentation, basée sur le mécanisme des services à la personne et coordonnée par l'Agence nationale des services à la personne (ANSP), afin de créer une offre globale "matériel, connexion, formation" à destination des seniors. …/… Lancer une campagne plurimédias pour favoriser la confiance et les usages des TIC auprès des seniors ».* Mais, pour être efficace, ces mesures doivent bien prendre en compte les spécificités des PA en termes d'acculturation technique. En effet, différentes études (Sperandio *et al*, 1997) montrent qu'il n'y a pas de déterminisme générationnel quant à l'adoption de la technologie, si ce n'est les diverses dégradations et incapacités qui peuvent rendre difficile l'accès au système. Ceci dit, d'autres facteurs d'usage (d'utilité, d'utilisabilité, d'accessibilité et d'acceptabilité) peuvent également contrarier l'appropriation des technologies à destination des PA. Pour mieux identifier ces facteurs, nous proposons d'appliquer un modèle d'acceptation technologique TAM (*Technology acceptance model*) étendu (Hamner et Qazi, 2008) des PA du 3eme âge et du 4eme âge. Ce modèle est présenté dans la figure 1.

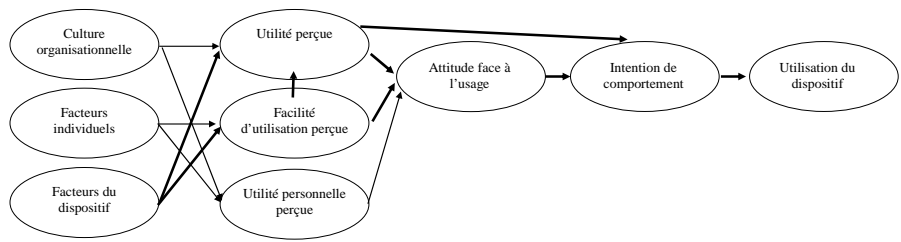

*Figure 1 : TAM étendu (Hamner et Qazi, 2008)*

Nous avons identifié certaines caractéristiques parmi ces facteurs à partir de la littérature. Elles concernent les facteurs individuels, les facteurs du dispositif, la culture organisationnelle et l'usage du dispositif. Nous les avons complétées et avons renseigné les autres facteurs (en particulier concernant la perception de l'utilité, de la facilité d'utilisation perçue, attitude face à l'usage et intention de comportement) à partir d'une étude sur le terrain auprès de PA du 4eme âge.

**TAM et PA du 3ème âge**

Dans le cas de l'usage des TIC chez les PA du 3eme âge, les facteurs qui relèvent de la culture organisationnelle correspondent à l'environnement social qui joue à la fois le rôle de tuteur pédagogique ou personne ressource en cas de problème (Karavidas *et al.,* 2005) ou un rôle de miroir qui valorise l'ego de la PA par son usage (Dickinson and Gregor, 2006) (Caradec, 2004). Concernant l'entraînement et la formation, Morrell (2005) mentionne qu'il n'y a pas de méthode optimale identifiée mais quelques critères (par exemple une formation longue, plutôt en binôme ou en petits groupes et avec des séquences d'apprentissage non complètement dirigées). Il indique de plus que ces formations permettent aux PA du 3eme et 4ème âge d'acquérir et maintenir des compétences sur du long terme. Les facteurs ergonomiques spécifiques au dispositif concernent l'accessibilité et l'utilisabilité des interfaces (Czaja and Lee, 2007) (Morrell, 2005) (Osman, 2005) (Dickinson *et al.,* 2005) qui doivent être adapté aux déficiences psychomotricielles ou sensitives (ouïe, vue, cognition) des PA. Par exemple, Morrell (2005) propose un guide de conception de site web accessible pour les personnes âgée et indique les caractéristiques typographiques adaptées à des déficiences de vision et des caractéristiques d'organisation des pages web adaptées à des déficiences cognitives (texte en voie active ne nécessitant pas d'inférence pour sa compréhension, langage simple, texte organisé en petites sections, etc.) et mentionne l'importance d'une navigation simplifiée. Dickinson (Dickinson et al., 2005) présente les caractéristiques d'utilisabilité d'un outil de messagerie. Un autre facteur individuel concerne l'usage préalable d'un dispositif TIC qui peut impacter directement l'attitude de la PA, en particulier son niveau d'anxiété face au dispositif, mais aussi la connaissance de l'utilité qu'il peut en avoir (Aula, 2005) (Karavidas *et al.,* 2005) (Marquié et Baccarat, 2001). Ces études montrent effectivement une amélioration de la qualité de vie, du niveau d'indépendance et de la reconstruction du lien social (Karavidas *et al.,* 2005), d'autres plus critiques précisent que ce n'est pas l'usage effectif du dispositif qui en est la cause mais la stimulation sociale liée aux phases d'entraînements (Dickinson et Gregor, 2006).

Ces études concernent généralement des PA du 3eme âge et peu d'informations sont disponibles concernant les PA du 4eme âge. Dans la suite,

nous proposons, à partir de l'étude expérimentale décrite ci-dessous, de formaliser, selon le modèle TAM une dynamique d'acceptation technique pour une population de PA du 4eme âge en EHPAD (Etablissement d'Hébergement pour Personnes Agées Dépendantes).

**TAM et PA du 4eme âge : Observation expérimentale**

L'étude expérimentale a été réalisée dans le cadre du projet MNESIS financé par le Ministère de la Recherche (appel « Usages de l'Internet »). Notre objectif était d'étudier les répercussions cognitives et psychosociales liées à l'usage d'un l'environnement technologique, Activital™, (Activital, 2008) sur une population de personnes très âgée résidant dans 7 EHPAD (Etablissement d'Hébergement pour Personnes Agées Dépendantes). Activital propose trois activités complémentaires (voir figures 2, 3, 4) : un ensemble de *jeux cognitifs*, un outil de *rédaction de journal de résidence* pour développer la créativité et un outil de *messagerie électronique* simplifié pour favoriser les liens sociaux et la communication.

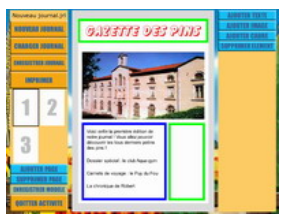  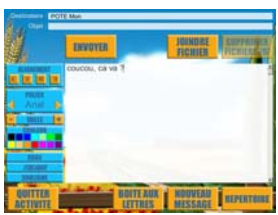  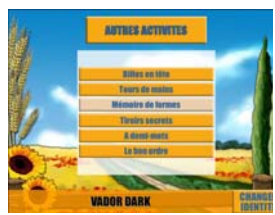

*Figure 2 : outil de rédaction de journaux*   *Figure 3 : outil de messagerie électronique*   *Figure 4 : jeux cognitifs à difficulté variable*

Un groupe de résidents a été formé à Activital. Il pratiquait des séances de formation à raison de 2 à 3 séances par semaine (durée de 30 à 45 minutes selon les EHPAD) sur une période de 6 mois de novembre 2005 à avril 2006. Ces sessions étaient assurées par les animateurs de chaque résidence qui avaient reçues pour l'occasion une formation spécifique de la part de la société éditrice du logiciel. Les autres jours, les ordinateurs restaient en libre service pour permettre aux PA de développer leur pratique. Nous avons observé cette expérimentation.

**Méthodologie d'analyse**

*Echantillon*

Sur une population potentielle de 45 personnes se déclarant intéressées pour être formées au dispositif et participer à l'étude, nous avons opéré une sélection pour satisfaire aux contraintes du protocole d'observation. Nous n'avons pas choisi de prendre les résidents ayant des déficiences cognitives (identifiées par un score inférieur à 25 au test "Mini Mental State2" (MMS) de Folstein (Derouesné *et Al.*, 1999). Nous avons préféré ne pas retenir non plus les personnes ayant des difficultés motrices et perceptives empêchant l'interaction avec l'environnement informatique, ou ayant des difficultés à s'exprimer et à rendre compte d'un vécu (personnel, social, psychologique…). Nous avons enfin privilégié les personnes ayant un entourage (familial ou amical) disposant d'une connexion internet pour permettre l'échange de messages électroniques. L'échantillon d'étude était donc de 27 personnes. 10 d'entre elles n'ont pas utilisé le dispositif informatisé mais des jeux identiques au format papier pour permettre d'avoir un groupe test dans les analyse cognitives (Tarpin-Bernard *et Al*, 2006). Sur les 17 personnes ayant suivi les ateliers informatiques, il y a eu 3 départs (retour au domicile ou décès). Nous considèrerons donc un échantillon stable de 14 participants sur toute la durée de l'étude, nous l'appellerons groupe MNESIS. Le tableau 1 expose les principales caractéristiques de cet échantillon.

| Age moyen | Genre | Situation matrimoniale | Profession antérieure | Descendance (moyenne) par résident | Ancienneté (moyenne) dans les résidences | Connaissance préalable de l'informatique |
|---|---|---|---|---|---|---|
| 84 ans (Ecart type 3,77) | 12 Femmes 2 Hommes | 11 veuves/veufs, 2 célibataires et 1 mariée | 7 employés adm., 4 fonctionnaires, 2 prof. indépendantes 1 sans profession | 3 enfants (~ 60 ans) 6 petits enfants (~30 ans) de 3 à 14 arrières petits enfants | 3 ans (Ecart type 2,23) | 4 par le travail (système à cartes perforées) 2 par la famille |

*Tableau 1 : Caractéristiques du groupe MNESIS*

*Méthode d'observation*

Divers recueils de données ont été effectués avant et après la formation à l'environnement technologique.

- des questionnaires auprès des 45 résidents initialement intéressés ainsi que quelques familles et personnel d'encadrement pour avoir une vue globale de la population du 4eme âge.

- des entretiens semi-directifs en début de parcours et à mi-parcours. Les entretiens initiaux permettait de compléter les questionnaires concernant le profil du résident, sa personnalité, ses relations sociales, son point de vue sur

---

[2] MMS : Ce test permet de repérer et de prévenir d'éventuelles dégradations cognitives

utilisation des TIC. Les entretiens à mi-parcours visaient à recueillir les premières réactions et impressions des résidents par rapport à l'outil, les usages effectifs, les apports, la facilité d'utilisation, la perception de l'entourage et le déroulement de la formation

- des observations sur le terrain de différentes natures : (1) observation de journée type (un observateur suit un résident et note tout ce qu'il peut faire du levé au couché), (2) observation de participation aux activités (les animateurs notent, pour chaque résident, quelles sont les activités proposées par la résidence qui sont choisies et l'attitude du résident dans ces activités), (3) observation du déroulement des séances de formation (un observateur renseigne, pour chaque problème rencontré par l'utilisateur le type de problème, la nature de l'intervention, qui l'aide et quelle fonctionnalité du logiciel pose problème).

- des recueils de traces d'usage. Pendant la formation nous avons procédé à l'enregistrement de toutes les traces d'interaction avec le clavier ou l'écran tactile.

Certaines techniques permettent de saisir la réalité des pratiques (*méthodes d'observation directes fondées sur les faits*), d'autres ont pour but de comprendre et d'expliciter ces pratiques (*méthodes d'entretien et de questionnaire fondées sur les représentations*). L'objectif était de parvenir à l'objectivation des données par la triangulation des méthodes et la mise en perspective de leurs résultats (Denzin, 1978, Vermersch 1994).

*Méthode d'analyse*

Les *questionnaires* auprès des résidents, familles et encadrement ont été traités avec Shinx et ont donné lieu à des rapports d'enquête quantitatifs (Oudart, 2005). Les *entretiens semi-directifs* ont été retranscrits et analysés manuellement pour saisir/affiner les points vues et contextes. Les citations extraites de ces entretiens sont présentées en italique dans l'analyse. La totalité des entretiens sont consultables dans (Oudart, 2005). Certaines *observations sur le terrain* (journée type et pointage de participation aux ateliers proposés par la résidence) ont été retranscrites en données quantitatives. Pour les *journées types* nous avons recodé chaque activité (toilette, lecture, jeux, repas, ….) selon son degré d'interaction sociale (exemple : 0 pour une activité passive individuelle et 5 pour une activité active collective). Les courbes des « journées types » ont permis d'analyser les résidents individuellement ou globalement comme dans la figure 4 proposée dans (Michel *et al.*, 2006a) qui présente une moyenne des niveaux d'interaction des résidents par heure, avant et après l'expérimentation. Les *pointages de participation aux ateliers* proposés par la résidence ont permis d'avoir une vision quantitative globale des activités choisies par les résidents et de leur attitude (active ou en retrait). Ainsi, le tableau 2 ci-dessous présente la moyenne mensuelle de participation aux activités pour tous les résidents comparativement à ceux du groupe MNESIS, et ce avant et après l'expérimentation. D'autres graphes, comptabilisant le nombre moyen de résident en attitude active ou en retraits sont présentés dans (Michel *et al.*, 2006a) (figure 6).

| | Activités d'expression et communication | Activités manuelles | Activités physiques | Activités socio-culturelles | Activité cognitives | Total |
|---|---|---|---|---|---|---|
| **Population totale DEBUT** | 1,97 | 0,77 | 0,40 | 2,36 | 0,61 | **6,12** |
| **Population totale FIN** | 2,29 | 0,91 | 0,37 | 3,79 | 1,17 | **8,53** |
| **Echantillon MNESIS DEBUT** | 2,60 | 0,92 | 0,38 | 2,58 | 1,68 | **8,17** |
| **Echantillon MNESIS FIN** | 5,26 | 1,47 | 0,86 | 4,33 | 1,44 | **13,36** |

*Tableau 2 : Moyenne mensuelle de participation aux activités en début et fin d'expérimentation pour l'ensemble des résidents et pour l'échantillon MNESIS*

Les enregistrements des *traces d'interaction* avec Activital ont été réalisés pour construire une base de connaissances sur les usages (Michel *et al.*, 2005) et ont été partiellement analysée par Esslimani (Esslimani, 2006). Elle a construit, à partir de tâches élémentaires (clic, frappe clavier et interaction de l'application) des tâches évoluées utilisées, de manière combinées, pour définir des profils d'usage.

**Résultats**

Nous présentons, dans les paragraphes suivants, les caractéristiques représentatives de chaque facteur du TAM étendu que nous avons pu identifier à partir des différentes méthodes d'observation.

*Facteurs individuels*

Le tableau 1 ci-dessus donne une première description socio-biographique de l'échantillon. Nous avons déjà présenté un bilan de l'existant socio-biographique complet (Oudart, 2005) ainsi qu'un bilan des pratiques sociales (Michel *et al.*, 2006a). Nous ne les représenterons pas ici et axerons l'analyse sur les facteurs individuels qui nous semblent pertinents pour expliquer le TAM. De part la constitution de l'échantillon, les résidents sont caractérisés par : une absence d'affection cognitive, d'handicap visuel ou auditif grave. Ils présentent une bonne capacité à verbaliser et disposent tous d'un entourage formé aux technologies informatiques. Le questionnaire et les entretiens semi-directif a montré qu'ils n'avaient aucune formation préalable aux applications proposées et à internet et souvent aucune manipulation d'un ordinateur ou autre dispositif technique mais une forte motivation malgré quelques craintes devant cet inconnu et une attente de changement dont les conséquences ne sont pas clairement identifiées. En effet, 82 % des résidents n'ont jamais utilisé d'environnements technologiques mais 65 % veulent essayer par curiosité, pour découvrir de nouvelles choses et 40,7% déclarent avoir déjà eu envie de se former à internet.

*Caractéristiques techniques du dispositif*

Comme nous l'avons déjà précisé, le logiciel Activital™ propose trois types d'activités : des *jeux cognitifs et ludiques* (ie, tour de Hanoï) ; un *outil de rédaction de journal de résidence* et un *outil de messagerie électronique*. Cet environnement prend en compte les difficultés d'usage de cette population à besoins spécifiques (contraintes perceptives, motrices…) en proposant une interface élaborée sur les principes d'accessibilité (WAI, 2007) et notamment les polices de caractères (taille, couleur), la structuration globale des écrans, la simplification du dialogue (vocabulaire et libellés des commandes) et le recours limité à la souris et au clavier (par la possibilité d'utiliser des écrans tactiles si besoin) comme le recommandent Morrell (2005) et Dickinson (Dickinson *et al.*, 2005).

Les entretiens semi-directifs effectués a mi-parcours ont toutefois identifié que l'usage du dispositif a pâti de nombreux dysfonctionnements techniques, surtout au début de l'expérimentation, en raison de bugs du logiciel mais aussi à cause de restrictions de sécurité pour l'accès au réseau imposés par la maison de retraite.

*Culture organisationnelle*

Nous considérons par culture organisationnelle le mode d'organisation (personnes de l'entourage, relations sociales, organisation professionnelle) et le mode de sensibilisation au TIC des résidents. Dans un EPAHD, dans la mesure où le lieu est fermé et le résident ne sort pas, les personnes de l'entourage de la PA sont l'encadrement (équipe médicale, direction, animateurs), la famille et les autres résidents. Concernant les relations sociales préalable à l'expérimentation, l'analyse des questionnaires et entretiens semi-directifs a permis de voir que depuis qu'ils sont en maison de retraite, la plupart des résidents ont l'impression être moins sociables et moins dynamiques. Les explications données sont soit personnelles (liées à l'âge, à des problèmes de santé) soit sociales (« *à cause des gens ici* », « *je m'ennuie* »). Le regard porté sur la population des résidents est d'ailleurs souvent très critique et dévalorisant *« Je ne suis pas à ma place car personne n'a toute sa tête »*. Quelques personnes âgées notent cependant que l'arrivée en résidence leur a permis d'être dans un environnement plus calme et serein, *« propice à découvrir de nouvelles choses »*. Les relations avec l'extérieur (famille et amis) se font principalement par téléphone (82 %) et en visite (74 %). De manière plus précise, les relations avec la famille sont jugées satisfaites mais la famille est souvent trop loin géographiquement ou sur-active ce qui empêche les visites. Les discussions tournent principalement autour de la famille, de la santé des personnes âgées et de sujets généraux (temps, faits divers…). La famille ne participe pas aux activités proposées par la résidence. Les relations avec l'encadrement sont considérées comme très satisfaite par 93% des résidents. Ils sécurisent les personnes âgées et leur donne l'impression d'une meilleure prise en charge. Les résidents gardent des relations avec d'anciens amis qui s'effectuent essentiellement par téléphone (67 %) et en visite (41 %). Les relations avec les autres résidents sont en revanche assez limitées (46% de l'échantillon) et plutôt superficielles.

Concernant la sensibilisation aux TIC, les animateurs et la famille ont joué un rôle clé dans la motivation et le passage à l'acte de la PA en suscitant plus ou moins volontairement une situation de challenge ou de « reconnaissance affective ». En effet, les entretiens semi-directifs ont relevé que les réticences initiales de certains résidents à participer aux ateliers informatiques étaient levées dès lors qu'ils cherchaient à faire plaisir à leurs animateurs ou familles en intégrant les groupes de formation. Ils voulaient également leur démontrer, qu'ils avaient la capacité à suivre cet apprentissage technologique. « *Je voulais montrer que l'on pouvait encore apprendre à nos âges* ». Pensant ne plus être attractif auprès de ses proches, il s'agissait alors d'étonner sa famille (enfants et petits enfants) et de susciter ainsi curiosité, intérêt et reconnaissance. La principale action de sensibilisation a été les séances de formation qui étaient principalement portée par les animateurs. Les relevés d'observation des séances ont montré leurs modes de réalisation : après avoir présenté quelques fonctionnalités du dispositif à manipuler lors de la séance (envoyer un mail) ou après avoir expliqué le fonctionnement du jeu, l'animateur restait généralement en retrait et assurait juste un soutien ponctuel. Il n'intervenait qu'en cas de dysfonctionnement majeur, ou suite à une demande du résident. La plupart de ces requêtes concernaient les jeux et étaient le plus souvent des demandes de soutien ou de réassurance psychologique (« *Je le fais bien* », « *C'est bien ça, non ?* » « *Je suis pas mal, hein ?* »). Si un second résident était présent, il proposait alors spontanément son aide. Les résidents ont mentionné systématiquement la gentillesse et le degré attention des animateurs à leur égard.

Pourtant, cette nouvelle activité n'a pas toujours été bien vécue par certains animateurs. La principale raison était l'absence d'assistance technique pour la gestion des ordinateurs (pannes informatiques entre autre) ou la manipulation des logiciels (gestion des bugs, mise en route, gestion des profils, etc.). L'équipe d'observation ou le directeur des animateurs a souvent eu ce rôle de soutien technique. Une autre raison a été le manque de reconnaissance de la Direction dans l'évolution de leur statut car à l'époque, ce type d'activité n'était pas inscrit dans les fiches de poste de ce personnel : c'était une tâche supplémentaire pour laquelle il n'avait pas été préparé. Plus globalement, la Direction a sous-estimé le rôle de ces acteurs dans le projet (d'autant plus que le *turn-over* des animateurs était important), et plus globalement sous-évalué les contraintes organisationnelles de la gestion d'un tel projet innovant : lieux d'implantation des ateliers informatiques, formation et motivation des animateurs, aléas techniques. Si la Direction a globalement bien communiqué auprès des familles et des participants à l'étude (sur les tenants et les aboutissants du projet), elle a en revanche peu discuté avec son personnel sur les apports et enjeux de ce projet ; ce qui a pu être perçu comme contrainte supplémentaire venant contrarier ou « sur-charger » leur activité. Les réticences et résistances étaient donc nombreuses. Par exemple, les réunions (ou tout autre forme de séance d'information) en grand nombre ont été difficiles à organiser, ce qui a engendrer des défauts d'information aux animateurs sur l'organisation des séances et le non respect des plannings (calendrier des séances, type d'activité messagerie-jeu-journal, durée des formations).

*Facilité d'utilisation perçue*

Dans le questionnaire, à la question, *« vous sentez-vous capable d'utiliser ces ordinateurs, ces techniques ou ces écrans ? »,* 29,6% pense que oui, 11,1% pense que non, 22,2% ne savent pas et 37% ne répondent pas. A la question ouverte complémentaire *« pourquoi »* beaucoup répondent *« j'ai encore toute ma tête ».* La plupart se sentent capable d'utiliser Internet (47%) et n'ont pas de crainte voire même sont plutôt enchantés à l'idée de l'utiliser : « *Je ne connais pas Internet mais je n'ai pas de craintes et si je retourne chez moi un jour je pourrais me servir de l'ordinateur »,* « *Je considère qu'il faut vivre avec l'évolution des vies et Internet en fait partie* ». Le fait que leur entourage ait une pratique de l'informatique ou d'Internet les a sensibilisés, familiarisés, désinhibés. Ils ont l'impression de baigner en quelque sorte dans cette ambiance technologique *« Je savais ce qu'était Internet par mon fils qui m'a déjà tout expliqué».* Cette stimulation passive a facilité l'acceptation de l'informatique et a abaissé la crainte même s'il a subsisté un sentiment d'incompétence dû à une faible estime d'eux-mêmes (Michel *et al.*, 2006b).

*Intention de comportement*

Les motivations et attentes des résidents par rapport au projet et aux TIC avant l'utilisation sont les suivantes : 65 % veulent essayer par curiosité, pour découvrir de nouvelles choses, 18% pour se prouver qu'ils sont encore capable d'apprendre et 18,5% pour rencontrer et échanger avec des personnes extérieures. Les intentions de comportement ne sont donc pas clairement identifiées.

*Utilité perçue*

L'utilité perçue des jeux est souvent mentionnée pour pallier les déficiences de mémoire. Concernant le mail et à terme internet, 47 % en attendent une ouverture sur le monde et 29 % une meilleure interaction avec l'entourage familiale et amicale, « *Je voulais envoyer des messages à mes enfants et petits enfants qui sont éparpillés dans toute la France et tous équipés d'ordinateurs »,* mais beaucoup de réponses concerne la satisfaction d'une curiosité «*je veux être au courant des évolutions »,* « *il faut vivre avec son temps »* ou bien pour saisir l'opportunité qui leur ait offerte sans avoir a priori d'utilité *« ça peut toujours servir ».* Aucun résident ne mentionne l'utilité du traitement de texte.

*Attitude face à l'usage*

Lors des entretiens semi-directifs, les résidents se sont déclarés plutôt contents de l'usage de l'informatique. Certains ont considéré la formation comme une nouvelle activité qui occupait la journée notamment en ce qui concernait les messages (pré-rédigé sur papier avant de les adresser par mail le jour de l'atelier) *« C'est une activité nouvelle qui me sort de mon quotidien ».* D'autres ont ressenti un bénéfice direct lié aux jeux, *« le fait de progresser dans les niveaux montre que l'on devient meilleur, que mon niveau intellectuel progresse… »* Ce bénéfice, quantifié par les utilisateurs via le score, comme le montre la citation, a été crucial dans l'usage, et nous avons vu que la demande d'être rassuré par les animateurs était forte pour ce type d'activité. Ils ont déclaré souhaiter progresser et approfondir leurs premières connaissances de l'informatique avant d'apprendre d'autres technologies (comme une caméra numérique par exemple). Suite à l'expérimentation, 2/3 des résidents ont exprimé le besoin d'avoir un ordinateur personnel : *« J'ai envie de me procurer un ordinateur dans ma chambre. Mon fils en a deux chez lui et j'espère qu'il m'en cédera un …/… Je pense que cela devient de plus en plus nécessaire dans ma vie. »..* Ils se sont déclarés particulièrement fiers d'eux : *« Je suis fier de ce que j'ai fait. De plus, j'ai l'impression que les jeux améliorent la mémoire, l'intelligence ». « Satisfaction personnelle car je suis assez fier de dire que je maîtrise l'informatique. Cela m'a apporté de la confiance et de la satisfaction»* – *« L'informatique ne m'a apporté que des choses positives ».* Les résidents mettent en avant l'innovation que représente ce projet dans leur vie de tous les jours au sein de la résidence. Il leur permet de connaître quelque chose d'actuel, de vivre avec son époque, de connaître quelque chose de nouveau (dont tout le monde parle) avant de mourir… Finalement ils ressentent le fait d'exister encore et d'être capable d'apprendre, d'évoluer, de progresser (et de ne plus régresser) *« C'est une saine occupation qui ne peut être que bénéfique à la connaissance (…) Il s'agit de ne pas mourir idiot » « C'est quelque chose de nouveau qui me stimule et m'amuse. Je peux apprendre beaucoup de choses nouvelles ».* Ils sont également fortement encouragés à continuer par leur famille.

Paradoxalement, l'usage de l'informatique fait naître un sentiment de finitude, de fin prochaine et inéluctable. Nombreuses étaient en effet les PA, se déclarant frustrées parce qu'elles n'auraient plus le temps de s'investir dans l'informatique alors que l'envie d'apprendre, de se former est de plus en plus forte *« c'est malheureusement trop tard, je suis trop vielle, trop âgée, je n'aurai plus le temps. (…) C'est plus difficile à mon âge, alors à quoi bon continuer ».* Cette expérience leur a fait aussi prendre conscience certaines limites et incapacités personnelles (motrices, cognitives, perceptives) dans l'accès aux technologies. *« C'est dur avec la souris car j'ai la main qui tremble. Je ne m'en étais pas rendu compte que cela pouvait être aussi gênant auparavant »* ou encore *« J'ai découvert que j'avais beaucoup de mal à lire* *sans lunette l'écran : j'avais mal à la tête et avais des troubles de la vision. J'ai donc dû prendre RDV avec l'ophtalmo. »*

Au travers de leurs discours, il ressort aussi que l'acquisition d'une connaissance informatique par les formations délivrées a conféré aux PA participant à l'étude un rôle d'expert au sein de la résidence, qui a créé une forte distinction, pour ne pas dire discrimination au sein de l'établissement entre : (i) ceux qui n'ont pas été sélectionnés pour le projet et qui montrent leur déception par un désintérêt de l'activité informatique et un relative indifférence envers les personnes qui y participent ; et (ii) ceux qui sont associés au projet Mnésis et qui ont eu l'impression de s'élever intellectuellement et « statutairement » par rapport au reste des personnes de la résidence ; en tout cas d'avoir tout mis en œuvre pour ne pas régresser. Certains se considèrent d'ailleurs comme faisant partie des pionniers, voire d'une élite. Ils sont très critiques à l'encontre de tous ceux qui se laissent dépérir et qui ne font pas l'effort de s'investir dans les divers ateliers proposés (dont l'informatique). Un système « paradoxant » -inhérent à la dynamique des groupes (Mucchielli, 1992)- s'est en quelque sorte développé : plus les participants ressentent un sentiment de confiance et d'estime de soi, plus ils construisent un système fondé sur la cohésion sociale du groupe ; et plus à l'inverse, ils développent un sentiment de dépréciation et de mépris à l'encontre des plus faibles. *« Les autres ne suivent pas, ils sont des boulets, trop lents. Je choisis mes activités pour ne pas me retrouver avec eux. Sinon, ça ne suit pas et je ne réussis pas ! »*

Enfin, nous avons pu constater que plus ils se formaient, plus ils devenaient exigeants. Si l'environnement était toujours jugé ergonomiquement satisfaisant, les temps de connexion devenaient trop lents, les services proposés pas assez variés (l'accès à Internet est souvent demandé et les jeux considérés comme rapidement complètement explorés) et le personnel perçu parfois comme incompétent car incapable de résoudre toutes leurs requêtes et aléas techniques.

*Usage*

La fonction de l'environnement technique privilégiée par les résidents est les jeux cognitifs. *« Cela m'a apporté de la confiance et de la satisfaction »*, *« cela entretient la mémoire, cela nous fait travailler »*, *« j'ai appris beaucoup de choses grâce au jeu de mots, je suis au niveau 2 c'est très stimulant »*. Ces jeux de stimulation les épanouissent sans les infantiliser. Cette activité étant survalorisée par les résidents, le niveau d'attente et d'exigence devient très élevé en terme de possibilités techniques et de niveau d'encadrement. Les tests relatifs aux incidences cognitives du système qui ont été conduits par le laboratoire EMC de Lyon 2 ont d'ailleurs montré qu'il n'y a eu ni dégradation, ni amélioration des personnes sur le plan cognitif, mais plutôt une stabilisation des processus supérieurs (raisonnement, mémoire…) (Tarpin-Bernard *et al.*, 2006).

Le mail a été utilisé de manière plus sporadique mais est bien identifié comme un palliatif à « l'enfermement » de la résidence et une bonne solution pour renforcer ou créer des liens sociaux (par la famille, des amis perdus de vue ou même avec le personnel). *« J'ai l'impression qu'on a reformé une famille beaucoup plus qu'avant grâce à l'internet ». « J'ai envoyé un mail à des amis américains que je ne vois plus. Cela va me permettre d'avoir des nouvelles plus fréquemment. Si je peux correspondre avec eux plus facilement, ce serait une bonne chose». « J'ai également reçu une pièce attachée : une photo c'était très gentil. La photo des enfants d'une infirmière ».* Les résidents considèrent qu'ils sont relativement autonomes dans la manipulation du mail ou des jeux. En revanche, ils expriment avoir encore des difficultés d'utilisation avec le journal pour saisir rapidement des textes, les structurer et faire des mises en formes. L'analyse des traces d'usage du mail et d'un jeu initié par Esslimani (2006) apporte quelques informations complémentaires. Nous les présentons plus a titre d'illustration de ce qu'il est possible de faire qu'à titre de résultat à proprement parlé car tous les participants n'ont pas été analysés.

Comme nous l'avons déjà précisé, elle a construit des profils d'usage en combinant les taches évoluées identifiées à partir des clics et frappe clavier. Par exemple, pour l'usage des jeux, Esslimani a défini les profils suivants : *consciencieux* (correspond à la réalisation de toutes les phases du jeu dans un ordre séquentiel et suivant les instructions préétablies de ce jeu), *zappeur* (correspond à un processus de réalisation marqué par des retours dans les phases antérieures du jeu, ou par une répétition de phases du jeu), *incertain* (correspond à une hésitation et une incertitude dans la façon de réaliser le jeu , par exemple : le fait d'entamer le jeu et quitter subitement). Pour évaluer les changements dans la manipulation technique et l'apprentissage autour du jeu, Esslimani a extrait 4 indicateurs : le *nombre d'essais*, la *rapidité de réalisation du jeu*, le *choix du niveau de difficulté* fait par le résident, et le *score au jeu*. Le choix des niveaux de jeu donne une indication sur l'appréciation que le résident a de son apprentissage du jeu (et permet de déterminer si l'apprentissage est positif ou négatif) mais aussi, combiné avec le score, sur son degré de confiance envers l'outil. Les analyses ont montré que le profil de «zappeur » a été identifié dans la moitié des cas lors de la réalisation de l'activité «Hissons les couleurs». L'observation des autres variables (voir tableau 3) montre un apprentissage autour de ce jeu mitigé. Sur les 6 résidents, deux ont abandonnés et pas réessayé. Pour les autres l'apprentissage est positif. En effet, si l'on regarde le niveau choisi, tous sont globalement assez peu confiance en eux au début de l'expérimentation (niveau facile ou moyen choisi) et augmentent progressivement. Les scores sont parallèlement plutôt croissants.

| Utilisateur | Niveau choisi | Date | Durée de réalisation | Score |
|---|---|---|---|---|
| 1 | Moyen | 2005-04-22 | 203 s | 16% |
| | Difficile | 2005-07-01 | 439 s | 66% |
| 2 | Facile | 2005-04-21 | 51 s | 100% |
| | Difficile | 2005-04-22 | | |
| | Moyen | 2005-05-20 | 32 s | 33% |
| | Difficile | 2005-05-20 | 48 s | 50% |
| 3 | Facile | 2005-04-22 | 202 s | 100% |
| | Facile | 2005-04-22 | 114 s | 100% |
| | Facile | 2005-05-04 | 93 s | 100% |
| | Facile | 2005-05-04 | 101 s | 66% |
| | Facile | 2005-05-04 | 72 s | 100% |
| | Moyen | 2005-08-05 | 171 s | 100% |
| | Moyen | 2005-08-05 | 195 s | 100% |
| 4 | Moyen | 2005-04-22 | ---- | ---- |
| 5 | Difficile | 2005-04-22 | 107 s | |
| | Moyen | 2005-04-22 | 47s | |
| | Facile | 2005-05-03 | 63 s | 100% |
| | Moyen | 2005-05-03 | 224s | 100% |
| | Difficile | 2005-05-03 | 48 s | 100% |
| 6 | Moyen | 2005-04-22 | 90 s | ---- |

*Tableau 3 : Exemple de recueil d'usage pour l'activité jeu « Hissons les couleurs »*

De la même manière, pour appréhender les changements d'usage liés au mail, Esslimani a identifié, pour chaque session : la *durée de la rédaction du mail* (en seconde), le *nombre de caractères* à l'envoi du message et le *nombre de caractères frappés*, la *vitesse de rédaction* (en caractère par seconde), le *degré d'erreur* qui est le pourcentage de caractère frappés non conservé dans le texte final du message, les fonctions de messagerie utilisées, le *type de fonction* utilisée (par exemple : gestion du répertoire, rédaction message, jointure fichier, mise en page, lecture message, etc.).

Sur la base de 4 utilisateurs analysés, elle a identifié une vitesse de frappe homogène. Si l'on considère qu'un mot fait en moyenne 6 caractère, les PA on une vitesse de frappe maximum de 1 caractère par seconde soit approximativement 10 mots par minute (a titre de comparaison un adulte actif professionnel a une vitesse de frappe de 50-60 mots par minute). Ces valeurs ne sont pas surprenantes au regard de la population concernée. Le degré d'erreur est de 15-20% en général sauf pour un utilisateur qui a 50%. Les séquences d'action et les services utilisés sont assez variés mais sont, vu le peu de séances précédentes (3 séances), probablement réalisés à la demande de l'animateur pour l'apprentissage.

**Impact : incidences des technologies**

L'analyse de l'incidence des TIC (Michel *et al.,* 2006a) a permis d'identifier une *(re)construction psychosociale de la PA* caractérisée par : a) une revalorisation de la PA, de l'estime de soi, b) une stimulation cognitive de stabilisation des processus supérieurs (raisonnement, mémoire…), c) une stimulation sociale, d) de nouvelles pratiques sociales de collaboration. La *revalorisation* est liée à « l'image positive de soi » qui se dégage de l'usage des environnements et qui se traduit par un gain de confiance dans leurs actes et un gain d'autonomie et d'initiative. L'image de soi est aussi directement revalorisée par le sentiment d'exister via la maîtrise d'un dispositif a priori complexe (et perçu par certains comme inaccessible) et via aussi la reconnaissance que leur porte leur environnement social et familial. La revalorisation est enfin soutenue par la mobilisation individuelle vers et dans un collectif et par l'acquisition ou la remobilisation de « compétences », de connaissances à partir de la pratique (technique, d'organisation, dextérité sensori-motrice, rédactionnelle, etc. ). La *stimulation cognitive* est une conséquence de l'usage des jeux ou liée aux réflexions nécessaires en amont de l'écriture d'un message électronique ou d'un article du journal électronique, en terme de recherche d'information ou formalisation de la pensée. La revalorisation et la stimulation cognitive sont donc des incidences directes de l'usage. La *stimulation sociale* aussi est liée au dispositif mais pas au sens premier supposé. La communication n'est que partiellement médiée par le dispositif (moins de trois mails sont en moyenne envoyés ou reçus par séance), en revanche les conversations avec la famille se voient enrichies et diversifiés de thèmes liés à l'informatique. Concernant le dernier point, l'idée de *collaboration*, l'utilisation de l'informatique a fait naître une sorte de collectif, ou plutôt de groupe en formation, où les personnes se parlent, se découvrent, s'aident et s'entraident et où une certaine complicité s'installe.

Sur le plan de l'identité sociale et de la stimulation cognitive, on peut dire que, à l'instar des « gérontechnologies », le dispositif agit directement en réduisant les diverses dégradations sensori-motrices, psychologiques ou cognitives. Sur le plan social et collaboratif en revanche, le dispositif agit principalement comme un artéfact symbolique. Néanmoins, dans un cas, comme dans l'autre, il apporte des gains notables en termes de qualité de vie mais aussi de gain de sens et d'objectif de vie.

**Conclusion**

Notre recherche se proposait de décrire les modalités d'accès et les dynamiques de sensibilisation ou d'acculturation aux dispositifs TIC des séniors. L'état de l'art a montré une offre restreinte en terme d'accès et en termes de services qui soient adaptés et accessibles. En cas d'offre de service adapté, les études ont montrés que pour les PA du 3$^{ème}$ âge, il y avait généralement une acculturation c'est-à-dire un apprentissage réel non seulement des fonctionnalités mais aussi des règles d'usage. Concernant notre population du 4eme âge en institution, les résultats ont montré que la sensibilisation a été complètement réussie et qu'une acculturation lente commençait à se produire. Le modèle d'acceptation technologique construit est en effet caractérisé par plusieurs facteurs qui auraient pu freiner le processus : les facteurs individuels (pas de connaissance préalable, un grand âge, de la crainte), une absence de perception d'utilité du dispositif (sauf concernant les jeux) et une intention de comportement non formalisée. Néanmoins, les facteurs du dispositif sont favorables (dispositif correctement configuré, c'est-à-dire adaptés et accessibles pour ce profil spécifique d'usager) et l'organisation sociale ainsi que le service à la personne sont vécus comme des soutiens efficaces par les résidents. Ces deux paramètres sur-stimulent la facilité d'utilisation perçue ce qui rend l'attitude face à l'usage particulièrement positive.

En terme d'impact, l'environnement technologique, en particulier les jeux de stimulation cognitive mais aussi dans une moindre mesure la messagerie, ont apportés des capacités et un pouvoir d'agir. A ce titre, la stimulation intellectuelle et l'importance des relations interpersonnelles suscitées par l'usage ou la rencontre avec ces nouvelles technologies s'avèrent être de puissants mécanismes d'adaptation pour aider ces personnes âgées à retrouver une intégrité cognitive (stimulation, rééducation), physique (par l'ouverture virtuelle que les TIC apportent sur l'environnement), mais aussi psychologique (reconnaissance et valorisation) et sociale (maintien/accroissement du lien social et de l'intégration sociale). En effet, l'identité sociale de la personne âgée a évolué positivement et plus globalement, cette étude a montré le fort impact des dispositifs TIC adaptés et correctement présentés en termes de gains en qualité, sens et objectifs de vie pour cette population. Elle renforce l'idée que pour cette population, plus que des questions d'accès, il faut réfléchir, en parallèle à la définition de dispositifs adapté et à la création de lieux d'accueil, de sensibilisation, discussion et formation si l'on souhaite réduire la fracture sociale.

## Bibliographie